\documentclass[prb,aps,epsf,twocolumn,showpacs]{revtex4-1}
\usepackage{dcolumn}
\usepackage{bm}
\usepackage{epsfig}
\usepackage{latexsym}
\usepackage{amsmath}
\usepackage{amssymb}
\usepackage{color}
\usepackage{array}
\usepackage{ulem}
\usepackage{cancel}
\usepackage[utf8]{inputenc}
\let\oldmarginpar\marginpar
\renewcommand\marginpar[1]{\-\oldmarginpar[\raggedleft\footnotesize #1]%
{\raggedright\footnotesize #1}}

\newcommand{\be}{\begin{equation}}
\newcommand{\ee}{\end{equation}}
\newcommand{\bea}{\begin{eqnarray}}
\newcommand{\eea}{\end{eqnarray}}

\renewcommand{\vec}[1]{{\bf #1}}

\newcommand{\e}{\varepsilon}

\renewcommand{\>}{\rangle}
\renewcommand{\(}{\left(}
\renewcommand{\)}{\right)}
\renewcommand{\[}{\left[}
\renewcommand{\]}{\right]}
\renewcommand{\v}[1]{\mathbf{#1}} % \v -> vector (bf)

\renewcommand{\cite}[1]{[\onlinecite{#1}]}

\begin{document}
\title{Marginal Anderson localization and many body delocalization}
\author{Rahul Nandkishore$^1$}
\author{Andrew C. Potter$^2$}
\affiliation{$^1$Princeton Center for Theoretical Science,Princeton University, Princeton, New Jersey 08544, USA}
\affiliation{$^2$Department of Physics, University of California, Berkeley, CA 94720, USA}
\date{\today}
\begin{abstract}
We consider $d$ dimensional systems which are localized in the absence of interactions, but whose single particle (SP) localization length diverges near a discrete set of (single-particle) energies, with critical exponent $\nu$.  This class includes disordered systems with intrinsic- or symmetry-protected- topological bands, such as disordered integer quantum Hall insulators.  In the absence of interactions, such marginally localized systems exhibit anomalous properties intermediate between localized and extended including: vanishing DC conductivity but sub-diffusive dynamics, and fractal entanglement (an entanglement entropy with a scaling intermediate between area and volume law).  We investigate the stability of marginal localization in the presence of interactions, and argue that arbitrarily weak short range interactions trigger delocalization for partially filled bands at non-zero energy density if $\nu \ge 1/d$. We use the Harris/Chayes bound $\nu \ge 2/d$, to conclude that marginal localization is generically unstable in the presence of interactions. Our results suggest the impossibility of stabilizing quantized Hall conductance at non-zero energy density.
\end{abstract}
\maketitle

\section{Introduction}
Edge states of topological systems give rise to a striking set of coherent quantum phenomena.  For example, chiral edge states of two-dimensional quantum Hall systems lead to precisely quantized Hall conductance at zero temperature.  However in thermal systems with non-zero temperature, these edge-state properties are washed out due to thermal excitations that communicate between opposite edges of the system.  Such deleterious thermal effects can sometimes be mitigated by careful cooling and isolation.  Nevertheless, it is interesting and potentially useful to ask whether one can obtain systems with topological edge state properties that are immune to thermal washing out, circumventing the need for cooling. 

An intriguing possibility arises from the study of many-body localization (MBL) - a phenomenon whereby well-isolated quantum systems fail to thermalize due to the localization of excitations by strong disorder \cite{Anderson, Fleishman, agkl, Mirlin, BAA, pal, oganesyan, Prosen, VoskAltman, Imbrie, arcmp}.  In MBL systems, disorder can protect quantum coherence against such thermal degradation by localizing the offending excitations.  In this fashion certain coherent quantum phenomena including symmetry breaking and topological order may survive in MBL systems at non-zero energy density (the appropriate analog of finite temperature in non-thermal systems) \cite{LPQO, Pekker, Vosk, Bauer, Bahri, Chandran}.

This naturally raises the question of whether topological edge states can be protected by localization of the bulk states.  Here, we face a new complication not present in previous examples of localization protected quantum order: topological bands in fermionic systems cannot be completely localized by disorder. Rather, the non-trivial band-topology guarantees the existence of an extended single-particle (SP) bulk orbital at some critical SP energy $E_c$\cite{Halperin} (or more generally a discrete set of such extended orbitals). SP orbitals with energy $E$ near $E_c$ exhibit diverging SP localization length \cite{endnote}:
\begin{align} \xi(E) \sim \frac{1}{|E-E_c|^\nu} \label{eq:DivergingXi}\end{align}
where $\nu$ is an exponent whose value depends on the particular system under consideration (see Fig.1).  We dub such systems ``marginally localized."  
\begin{figure}[ttt]
\begin{center}
\includegraphics[width = 3.3in]{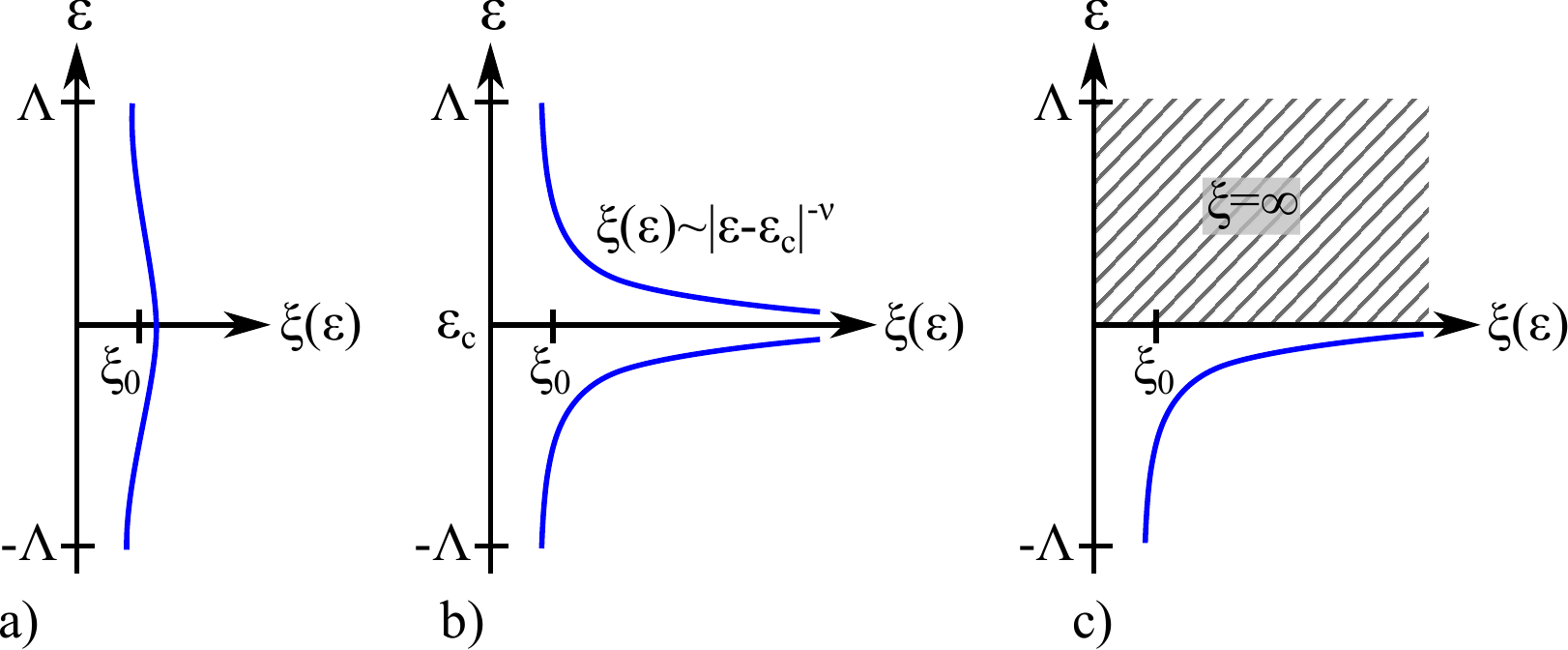}
\end{center}
\vspace{-.2in}
\caption{Schematic dependence of SP localization length $\xi$ on  single-particle energy $E$ for a fully-localized system (a), marginally localized system (b), and extended system with a mobility edge (c).
}
\vspace{-.2in}
\label{fig:DivergingLocLength}
\end{figure} 
Examples include: disordered integer quantum Hall systems \cite{QuantumHallPlateau,WannierObstruction} and chiral superconductors, intrinsically topological superconductors \cite{Motrunich}, and symmetry protected topological phases (e.g. topological insulators) with symmetry-preserving disorder \cite{Z2WannierObstruction}.  Here, the diverging localization length accompanies each change in the topological edge-state structure (e.g. the Chern-number changing quantum Hall plateau to plateau transitions). The presence of at least one extended state is guaranteed by the topological obstruction to constructing localized Wannier orbitals in a topological band \cite{WannierObstruction,Z2WannierObstruction}.  
In the vicinity of the critical energy, the SP localization length diverges as in Eq.~\ref{eq:DivergingXi}.  A diverging SP localization length can occur in non-topological contexts, such as 1D XX spin chains with random bond disorder (or equivalently, non-interacting fermion chains with random hopping). Such marginally localized systems have properties intermediate between localized and extended systems, as we shall discuss in Sec.II. 

A central issue is whether marginal localization persists away from the non-interacting limit.  In fully localized systems ($\xi(E)\leq \xi_0$ for all $E$) there is strong evidence that localization survives the addition of weak interactions even at non-zero energy density \cite{BAA, oganesyan, pal}.  In contrast, interactions are expected to delocalize a system whose single particle (SP) spectrum contains a mobility edge, as a generic excited state will contain a finite density of occupied extended orbitals that thermalize among themselves and then act as a bath for the localized orbitals. In the presence of a non-zero coupling to a bath, the many body wave functions become thermal, and localization is destroyed, even though signatures of proximity to localization may survive in the spectral functions of local operators \cite{ngh, jnb}. Even if extended states only arise in a narrow energy range of width $W$, the results of \cite{gn} suggest that this narrow band of extended states will be sufficient to destroy localization, for any non-zero $W$. However, for marginally localized phases, extended states arise only at a single energy $E_c$. In the language of \cite{gn}, these extended states constitute a bath with vanishing bandwidth, and it is not a priori clear whether this will be sufficient to destroy localization upon inclusion of interactions. 

In this paper, we give analytic arguments showing that localization is impossible for arbitrarily weak interaction strength if $\nu d > 1$, where $d$ is the spatial dimension. The breakdown of localization originates from a proliferation of multi-particle resonances between well separated single-particle orbitals. Notably, such breakdown occurs in disordered integer quantum Hall systems for which $\nu\approx 2.3$\cite{ChalkerCoddington,QuantumHallPlateau}. Moreover, in the generic case of systems that are governed by the Harris/Chayes bound, $\nu d \ge 2$, and which have a non-zero density of states (e.g. integer quantum Hall systems), our arguments preclude the existence of many body localization at finite energy density in partially filled bands where the localization length diverges at a discrete set of energies. 

Our arguments do not exclude many body localization in systems that evade the Harris bound due to symmetry reasons, and have $\nu < 1/d$. It also does not exclude localization in systems where the density of states vanishes sufficiently rapidly at the critical energy. For example, in a system where the density of states vanishes as $|E - E_c|^{\Upsilon}$, the breakdown of localization only occurs if $\nu d> 1+\Upsilon$. Finally, the Harris/Chayes bound applies to the critical exponent for the {\it mean} localization length, while our argument involves the critical exponent of the {\it typical} localization length, so there remains a possibility of marginal localization in systems where the mean and typical localization lengths diverge with very different exponents. 

This paper is structured as follows. In Section~\ref{sec:Properties} we review the properties of marginally localized phases in the absence of interactions. In Section~\ref{sec:Model} we introduce the model of interest, and the basic approach. In Section IV we derive a condition for the perturbative instability of marginal localization. The analysis in Section IV initially assumes that the wave functions are uniform on length scales less than a localization length, and decay exponentially on longer length scales. This is an oversimplification, since the wave functions close to criticality have multifractal character on length scales short compared to the localization length. In Section IV-H we take this multifractality into account, and demonstrate that it does not alter our essential results. In Section V we discuss the consequences of our results, and use the Harris/Chayes argument \cite{Harris, Chayes} to rule out stability of marginal localization in a wide class of models, including the integer quantum Hall effect. We also discuss the possibility of perturbatively stable marginal localization in systems that exploit certain loopholes in our argument, which we discuss. 

\section{Properties of Marginally Localized Phases \label{sec:Properties}}
Before analyzing the effects of interactions, we begin by reviewing the properties of marginally localized phases in the absence of interactions.  In this, and all that follows, we consider the case of only a single critical energy, $\e_c$ and shift our SP energy scale such that $\e_c=0$.  Moreover, we suppose that the SP density of states is regular and non-zero in the vicinity of $\e_c$. We work at partial filling in the band at a non-zero energy density (i.e. we are not discussing ground states or low lying excited states). In the quantum Hall context, this amounts to working at a non-zero temperature, but a temperature that is still much less than the cyclotron frequency (allowing us to focus on a single Landau sub band). 

\subsection{Subdiffusive relaxation}
Marginally localized states show anomalous subdiffusive dynamics, as is known from previous studies of dynamics in disordered quantum Hall systems\cite{Sinova,Sandler,Moore}.  These results are reproduced by the following simple argument. Consider the relaxation of a wave packet $\Psi$ initially localized in the vicinity of $\v{R}(t=0)=0$. The wave packet can be written in the form $|\Psi(t=0)\> = \sum_{\alpha} w_{\alpha} |\psi_{\alpha} \> $, where $|\psi_{\alpha}\rangle$ are the single particle states with energy $E_\alpha$ and amplitude $\varphi_{\alpha}(\vec{r})$ at a position $\vec{r}$. Meanwhile, $w_{\alpha}$ are appropriate complex amplitudes that produce a localized wave packet. The wave functions $\varphi_{\alpha}$ are all localized, with energy dependent localization lengths, $\xi(E_{\alpha})$ as in Eq.~\ref{eq:DivergingXi}.
This wave packet has the time evolution $ |\Psi(t)\> = \sum_\alpha w_\alpha |\psi_\alpha \rangle e^{-iE_\alpha t} $. We can now straightforwardly evaluate the spatial spreading of the wave packet by examining
\begin{align} R^2(t) &= \int d^dr \sum_{\alpha,\beta} e^{i(E_\alpha-E_\beta)t} w_\alpha^*w_\beta \varphi^*_\alpha(r)r^2\varphi_\beta(r)
\nonumber\\
&\approx \int d^dr \sum_{\alpha}  |w_\alpha|^2 \varphi^*_\alpha(r)r^2\varphi_\beta(r)\nonumber\\
&\approx \sum_{\alpha}|w_\alpha|^2\sum_{\beta:|E_{\alpha\beta}|<1/t} \min\(D_{\alpha}t,\xi_\alpha^2\)
\nonumber
\end{align}
where the second line is valid at long times when the phases between states with $|E_{\alpha\beta}|\equiv|E_\alpha-E_\beta|\gg 1/t$ have effectively randomized, and in going to the third line we have assumed that the spreading in each energy interval is diffusive up to the localization length at that energy, and have allowed for the possibility that the diffusion ``constant" varies with energy. The diffusion constant at an energy $E$ may be estimated by setting the Thouless energy $D(E) / \xi(E)^2$ equal to the level spacing $\xi^{-d}(E)$ at that energy, giving $D(E) \sim \xi^{2-d}$. The long-time dynamics will turn out to be controlled by states near the critical energy $E_c$. In dimensions $d\neq 2$. the singular behavior of $\xi$ near the critical energy will translate into a singular behavior of $D(E)$. However, in two dimensions (the case relevant for quantum Hall), the singularity in $\xi$ will not translate into a singularity in $D$, and thus we may replace $D(E)$ by a constant value $D_0$, the diffusion constant for the states near the critical energy. Moreover, the detailed initial initial structure of the wave-packet is irrelevant for the long-time asymptotics, which are captured by simply taking $|w_\alpha|$ to be roughly equal amplitude for all energies $E_\alpha$ within a bandwidth $\Lambda$ of $E_c$.  Under these assumptions, the spreading of the wave-packet follows:
\begin{align}
R^2(t) = \int_0^{E(t)} \frac{dE}{\Lambda}D_0t+\int_{E(t)}^\Lambda \frac{dE}{\Lambda}\xi^2(E)
\label{eq:R2}
\end{align}
In the fourth line we have defined the integration limit $E(t)$ through $\xi^2[E(t)] \approx \xi_0^2\(\frac{E(t)}{\Lambda}\)^{-2\nu} = D_0t$, i.e. 
$E(t)= \Lambda \(\frac{D_0t}{\xi_0^2}\)^{-1/2\nu}$. The behavior is qualitatively different for $\nu < 1/2$ and for $\nu > 1/2$. However, provided the Harris/Chayes argument\cite{Harris, Chayes} applies (see discussion in Appendix),  $\nu \ge 1$, and only the latter case is relevant. Thus, the dominant contribution comes from low-energy states with diverging $\xi(E)$, i.e. from the 1st term in Eq.~\ref{eq:R2} and from the lower end with $E\gtrsim E(t)$ in the 2nd term of Eq.~\ref{eq:R2}, and we obtain
\begin{align} R^2(t) \sim t^{\alpha}, \quad \alpha = 1-\frac{1}{2 \nu} \label{eq:Subdiffusion}\end{align}
Thus the dynamics are characterized by the dynamical exponent $\alpha = 1 - \frac{1}{2\nu}$ intermediate between diffusive ($\alpha=1$) and localized ($\alpha=0$) scaling, and taking the numerical value $\alpha \approx 0.78$ in the case of integer quantum Hall.  

\subsection{Scale dependent conductivity}
The sub diffusive spreading Eq.~\ref{eq:Subdiffusion} can be interpreted as a length-scale dependent diffusion ``constant": $D(L)\sim L^{-2/(2\nu-1)}$.
Using the usual Einstein relation between diffusion and conductivity, these considerations predict scale dependent DC conductivity
\begin{align} \sigma(L)\sim L^{-2/(2\nu-1)} \label{eq:DCConductivity}\end{align}
which vanishes in the the thermodynamic limit.  Again we see scaling intermediate between the constant conductivity $\sigma_\text{ex}\sim \text{const}$ of an extended system and the exponential scaling $\sigma_\text{loc}(L)\sim e^{-L/\xi}$ of a localized system.  

\subsection{Fractal Entanglement Scaling}
Entanglement scaling provides insight into the character of many-body eigenstates, and provides a conceptually powerful means to distinguish extended and localized systems. The entanglement entropy of a spatial sub-region of linear size $R$ for an excited energy eigenstate in a localized system comes only from states within a localization length, $\xi_0$, of the boundary and exhibits boundary-law scaling $S(R)\sim \xi_0 R^{d-1}$.  Here $d$ is the number of spatial dimensions.  In extended systems, such boundary law scaling occurs only for the ground-states, whereas excited states typically display volume law scaling $S(R)\sim R^d$.  In this section we analyze the spatial scaling of entanglement entropy in excited states of a marginally localized system, which has not been considered in previous studies.

For a marginally localized system, a fraction $f(R)\sim R^{-1/\nu}$ states are extended across the sub-region of size $R$ (i.e. have $\xi(\e)>R$), and make a volume-law type contribution to entanglement $f(R)R^d\sim R^{d-1/\nu}$.    For $\nu>1$, this dominates the area-law contribution from well-localized states leading to a scaling:
\begin{align} S(R)\sim R^{d-1/\nu} \label{eq:entanglement} \end{align}
intermediate between volume law and area-law, which we dub ``fractal" entanglement scaling as it scales like the boundary of a fractal sub-region of a localized phase with fractional dimension $d-1/\nu$.

Marginally localized systems exhibit fractal entanglement scaling so long as $\nu > 1$. For $\nu < 1$, the entanglement entropy becomes effectively area law. When applicable, a Harris-type criterion (see App.~\ref{app:Harris}) $\nu\ge2/d$ ensures that the fraction of modes which are extended on the length scale $R$ dominate the entanglement entropy in all systems with $d <2$. Integer quantum Hall systems have $\nu\approx 2.3$, and will exhibit fractal entanglement scaling.  However Harris/Chayes arguments do not rule out that some marginally localized systems exhibit conventional boundary-law scaling in spatial dimensions $d \ge 2$.

Thus, we see that in dynamics, response properties, and the scaling of entanglement entropy, marginally localized non-interacting systems display properties that are intermediate between traditional localized systems and delocalized systems. Next, we investigate whether this intermediate behavior survives in the presence of interactions. 

\section{Model and approach \label{sec:Model}}
The Hamiltonian of interest takes the form $ \hat H = \hat H_0 + \hat V$, where $H_0$ is the Hamiltonian of the disordered non-interacting fermion system (for example Landau levels with disorder). This non-interacting Hamiltonian has single-particle eigenmodes of energy $E_\alpha$ created by operators $\psi^\dagger_\alpha = \int d^dr \varphi^*_\alpha(r)c^\dagger(r)$, where $c^\dagger(r)$ creates an electron at position $r$. Meanwhile, the interaction term takes the form 
\begin{align}
\hat V = 
V \sum_{\alpha, \beta, \gamma, \delta} \lambda_{\alpha \beta \gamma \delta} \psi^\dagger_{\alpha}\psi^\dagger_{\beta} \psi_{\gamma}\psi_\delta \label{eq: V}
\end{align}
 This represents a short range interaction of strength $V$, where $\lambda_{\alpha\beta\gamma\delta} = \int d^dr \varphi_\alpha^*(r)\varphi_\beta^*(r)\varphi_\gamma(r)\varphi_\delta(r)$ are the matrix elements of the interaction in the eigen-basis of $H_0$.  We consider the simplest case (illustrated in Fig.~\ref{fig:DivergingLocLength}b) where the SP localization length diverges only at one SP energy, $E_c$, and choose our zero of energy to set $E_c\equiv 0$.  The localization lengths for energies close to zero then diverge as $\xi(E) \sim |E|^{-\nu}$, such that at any length scale $L$, a non-zero fraction $\sim L^{-1/\nu}$ of the states appear extended. We take the SP density of states to be a non-zero constant in the vicinity of $E=0$.  For generality we will keep the spatial dimensionality $d$ and the localization-length exponent $\nu$ arbitrary, for example $d=2$ and $\nu \approx 2.3$ for the disordered integer quantum Hall plateau \cite{QuantumHallPlateau}. 

We now ask whether inclusion of the interaction causes a breakdown of localization. We begin by noting that if we attempt to construct the many body eigenstates perturbatively in weak interactions, then perturbation theory for the eigenstates will break down already at $O(V)$ for states involving SP orbitals sufficiently close to the mobility edge, irrespective of the value of $\nu$. However, the fact that the eigenstates cannot be perturbatively constructed does not imply that localization itself is unstable - to address the stability of localization we need a more sophisticated diagnostic. 

It has been pointed out that fully many-body localized systems are locally integrable in the sense that all eigenstates can be completely specified by a set of exponentially well-localized conserved, commuting quantities or ``integrals of motion"\cite{Serbyn,HuseIntegrability}. For non-interacting systems the integrals of motion are simply the occupation numbers of single-particle orbitals.  In an MBL phase, these integrals of motion are dressed by interactions but (apart from a exponentially small fraction of rare resonances\cite{Imbrie}) retain their localized character. In this spirit, one could take the defining property of marginally localized systems to be the existence of a (nearly) complete set of algebraically well-localized integrals of motion.  Here, algebraically well-localized means that the fraction of integrals of motion with appreciable support on scale larger than $R$ decays as a power law in $R$, (unlike traditional MBL systems where it decays exponentially in $R$, and also unlike a delocalized system, where the integrals of motion are not localized), and the caveat ``nearly" allows for a possible set of non-percolating resonances.

We wish to investigate whether this picture of marginal localization can be stable in the presence of interactions. To this end, we first introduce some notation. We use $|\Phi_E\rangle$ to denote a (many-body) eigenstate of $H_0$ at an energy $E$, labelled by a particular set of marginally localized integrals of motion, in the sense described above. We use $|\Psi_E\rangle$ to denote a (many-body) eigenstate of $\hat H_0 + \hat V$ at an energy $E$. We now introduce the $\hat T$ matrix of the problem by $\hat V |\Psi_E \rangle = \hat T |\Phi_E \rangle$. The matrix elements of $\hat T$ between SP states with energy $E$ and spatial separation $R$ gives the effective matrix element for long range hopping. It is well known that the effective hopping matrix element must fall off with distance faster than $1/R^d$, else localization is impossible \cite{Anderson, Levitov2}. Similarly, the matrix elements of $\hat T$ between two particle states with the particles separated by a distance $R$ gives the effective matrix element for a long range two-body interaction, the matrix elements between three particle states give the effective matrix element for a long range three-body interaction etc. Again, there are conditions on how fast these matrix elements must fall off with $R$ in order for localization to be stable \cite{Fleishman, Burin, Yao}. Our main strategy in the present work will be to determine how the matrix elements of $\hat T$ fall off with distance $R$ between the particles involved. Hence, we will obtain a condition on $\nu$ for breakdown of localization.

 The $\hat T$ matrix is related to $\hat V$ by the Dyson equation \cite{Coopernotes}
\begin{equation}
\hat T = \hat V + \hat V \frac{1}{E - \hat H_0 + i 0} \hat T.
\end{equation}
This expression fully specifies the $\hat T$ matrix for states at a particular energy $E$. We note that even though $\hat V$ is short range, $\hat T$ can have long range matrix elements because $\hat H_0$ has eigenstates with arbitrarily long range matrix elements. Within perturbation theory in small $\hat V$, we can approximate the $\hat T$ matrix by the Born series
\begin{eqnarray}
\hat T &=& \hat V + \hat V \hat G_0(E) \hat V + \hat V \hat G_0(E) \hat V \hat G_0(E) \hat V + ... \label{Born}\\
 \hat G_0(E) &=& \frac{1}{E - \hat H_0 + i0} \nonumber
\end{eqnarray}
This perturbative expression can be represented in standard diagrammatic perturbation theory, where $G_0$ is represented by a (directed) line, and $V$ is represented by a vertex with two lines going in and two coming out. The diagrammatic representation of the $\hat T$ matrix contains only connected diagrams. 

\section{Delocalization is inevitable if $d\nu > 1$}
We use $T_{A,B}$ to denote the matrix element of $\hat T$ between incoming state $A$ and outgoing state $B$. If $T_{A,B}$ exceeds the level spacing between states $A$ and $B$ then these two states are `in resonance,' and the occupation number of state $A$ cannot be an integral of motion. However, any time two states are in resonance, we can define new states $A'$ and $B'$ which are orthogonal linear combinations of $A$ and $B$, such that $T_{A',B'} = 0$. The occupation number of $A'$ and $B'$ can then be an integral of motion (at least if the states are not resonant with any third state $C$). However, the new integrals of motion will have support wherever $A$ and $B$ had support. We see therefore that the spatial structure of resonances is crucial. If resonances only occur between states that have support at nearby points in real space, then we may be able to define new integrals of motion which are still localized, if the local resonances do not percolate \cite{Levitov2}. In contrast, if resonances occur on all length scales, then the integrals of motion will not be localized (even algebraically), and so the system will not be able to support marginal MBL. 

The question we are interested in asking is thus the following: if $A$ is an eigenstate of $H_0$ with localization length $\xi_A$, then what is the probability that $A$ is `in resonance' with another state $B$ a distance $R$ apart, where $R \gg \max(\xi_A, \xi_B)$ (the situation when $R < \max(\xi_A, \xi_B)$ constitutes a local resonance, which just redefines the localized integrals of motion). If the probability of having a resonance at a distance greater than $R$ goes to zero as $R\rightarrow \infty$, then we will be able to define algebraically localized integrals of motion with arbitrary precision, in the sense that all but a fraction $\epsilon$ of the integrals of motion will be localized on a length scale $R_c(\epsilon)$, where $R_c(\epsilon)$ is set by the condition that the probability of having a resonance at a length scale greater than $R_c$ is equal to $\epsilon$. In contrast, if the probability of having a resonance at a length scale greater than $R$ does not go to zero as $R \rightarrow \infty$, then we will not be able to define algebraically localized integrals of motion with arbitrary precision, and the system cannot support marginal MBL. 

We thus want to calculate how the matrix elements of $\hat T$ scale with distance. We recall that the $\hat T$ matrix is given by the Born series (\ref{Born}). The first order term $\hat V$ is purely short range. Long range terms first appear at order ($V^2$). We now examine the how the matrix elements of $\hat T$ fall off with distance at $O(V^2)$. 

\begin{figure}[t]
\includegraphics[width = 1.5in]{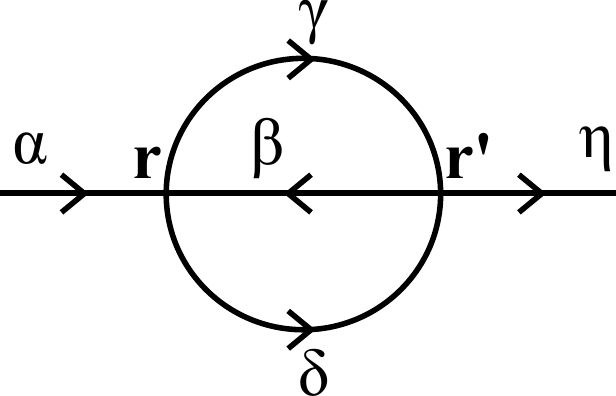}
\caption{Diagrammatic representation of mediated hopping, to lowest order}
\label{fig:LowOrderHopping}
\end{figure}

\subsection{Hopping resonances}
We begin by evaluating the matrix elements of $\hat T$ between well localized SP orbitals $\psi_{\alpha}$ and $\psi_{\eta}$ at a separation $R \gg \max(\xi_{\alpha}, \xi_{\eta})$. This can be viewed as the effective matrix element for hopping at a range $R$. 
The mediated hopping process at order $V^2$ is illustrated in Fig. 1. Thus, we have 
\begin{widetext}
\begin{align}
T_{\alpha , \eta}(R) &= \sum_{\beta \gamma \delta} \frac{V^2\lambda_{\alpha\beta\gamma\delta}\lambda_{\gamma\delta\beta\eta}
}{(E_{\alpha} + E_{\beta}) - (E_{\gamma} + E_{\delta})}
\approx \frac{V^2}{\xi^{d/2}_{\alpha} \xi^{d/2}_{\eta}} \sum_{\beta \gamma \delta} \frac{\exp\left[-R\left(\frac{1}{\xi_{\beta}} + \frac{1}{\xi_{\gamma}} +\frac{1}{\xi_{\delta}}\right)\right]}{\left(\xi_{\beta}^{d} \xi_{\gamma}^d \xi_{\delta}^d\right)(E_{\alpha} + E_{\beta} - E_{\gamma} - E_{\delta})}  \exp\left(-i \phi_{\beta \gamma \delta}\right)
\nonumber\\
&= V^2 |E^{\nu d/2}_{\alpha} E^{\nu d/2}_{\eta}| \sum_{\beta \gamma \delta} \Pi_{\beta \gamma \delta }(R)
\end{align}
where the sum ranges over intermediate states, $(\alpha,\beta,\gamma)$. Since orbitals $\alpha,\eta$ are well localized we have replaced: $\lambda_{\alpha\beta\gamma\delta}\approx \varphi^*_\alpha(0)\varphi^*_\beta(0)\varphi_\gamma(0)\varphi_\delta(0)$, and $\lambda_{\alpha\beta\gamma\delta}\approx \varphi^*_\lambda(R)\varphi^*_\delta(R)\varphi_\beta(R)\varphi_\eta(R)$. We approximate each of the wave-functions by the simplistic localized form: $|\varphi_\alpha(r)| \approx \frac{1}{\xi_\alpha^{d/2}}e^{-|r-r_\alpha|}$ (and similarly for $\beta,\lambda,\dots$). This assumption ignores the multi-fractal character of the SP orbitals with $\e$ close to $\e_c$, but is expected to reproduce the typical behavior. The multi-fractal character of the wave functions will be taken into account in Sec. V. For convenience we have defined:
\begin{eqnarray}
\Pi_{\beta \gamma \delta} (R) =|E_{\beta} E_{\gamma} E_{\delta}|^{d\nu} \frac{\exp\left[-R\left(|E_{\beta}|^{\nu} + |E_{\gamma}|^{\nu} + |E_{\delta}|^{\nu} \right)\right]}{(E_{\alpha} + E_{\beta} - E_{\gamma} - E_{\delta})} \exp \left(-i \phi_{\beta \gamma \delta}\right).
\end{eqnarray}
\end{widetext}

Each matrix element contains an overall phase, $\phi_{\beta \gamma \delta}$, and the phases are essentially uncorrelated for different choices of $\beta \gamma \delta$.  Hence the sum over intermediate states acculumates random amplitude and phase characteristic of a random walk in the complex plane of characteristic step size $\approx |\Pi_{\beta,\gamma,\delta}(R)|$.  The typical RMS modulus of the effective hopping is then:
\begin{eqnarray}
|T_{\alpha, \eta}(R)| &=& V^2 |E_{\alpha} E_{\eta}|^{\nu d/2} \sqrt{\sum_{\beta \gamma \delta \beta' \gamma' \delta'} \Pi^*_{\beta \gamma \delta }(R) \Pi_{\beta' \gamma' \delta'}(R) }, \nonumber\\
&=& V^2 |E_{\alpha} E_{\eta}|^{\nu d/2} \sqrt{\sum_{\beta \gamma \delta} |\Pi_{\beta \gamma \delta }(R)|^2},
\end{eqnarray}
where we have assumed that the `off diagonal' terms come with random phases and cancel out. 

We can trade the sum over discrete states for a continuous integral by identifying the appropriate density of states for intermediate energies $E_{\alpha,\beta,\delta}$. Recall that the localization length scales as $|E|^{-\nu}$. Thus there are $\approx E^{-d\nu} dE$ states with appreciable overlap with states $\alpha$ and $\eta$ in an energy interval $\[E,E+dE\]$. This implies that we should take $\sum_{\beta} \rightarrow \int dE_{\beta} |E_{\beta}^{-\nu d}|$. 
With the above rule for translating sums to integrals, we get 
\begin{widetext}
\begin{eqnarray}
\langle |T_{\alpha ,\eta}(R)|\rangle =  V^2 |E_{\alpha} E_{\eta}|^{\nu d/2} \sqrt{\int dE_{\beta} dE_{\gamma} dE_{\delta} |E_{\beta} E_{\gamma} E_{\delta}|^{d\nu} \frac{\exp\left[-2R\left(E_{\beta}^{\nu} + E_{\gamma}^{\nu} + E_{\delta}^{\nu} \right)\right]}{(E_{\alpha} + E_{\beta} - E_{\gamma} - E_{\delta})^2}}\nonumber
\end{eqnarray}
\end{widetext}
The exponential scaling of the wave-function amplitudes provides a cutoff, restricting us to states with $|E|_{\beta, \gamma, \delta} < R^{-1/\nu}$. Meanwhile, if $E_{\beta, \gamma, \delta}$ are all close to zero (which must be the case given the exponential cutoff) then the denominator is $\approx E_{\alpha}^2$, so we obtain
\begin{eqnarray}
\frac{\langle|T_{\alpha, \eta}(R)|\rangle }{V^2}&=& |E^{\nu d -1}_{\alpha}| \sqrt{2\int_0^{R^{-1/\nu}} dE_{\beta} dE_{\gamma} dE_{\delta} |E_{\beta} E_{\gamma} E_{\delta}|^{d \nu}}\nonumber\\
&=&|E_{\alpha}|^{d \nu -1} R^{-\frac{3}{2}(d +1/\nu)}
\end{eqnarray}

 This matrix element should be compared to the level spacing between the states $\psi_{\alpha}$ and $\psi_{\eta}$. If the matrix element is greater than the level spacing, then the two states are in resonance i.e. a particle originally in state $\psi_{\alpha}$ can readily hop to $\psi_{\eta}$. If we are to have any hope of localization, then the probability of having a resonance at a length scale $R$ must go to zero as $R \rightarrow \infty$. 
 
We recall that the effective matrix elements for long range hopping (at $O(V^2)$) fall off as $R^{-\frac{3}{2}(d +1/\nu)}
$. The probability that we have resonances at a distance $R > R_c$ is then
\begin{equation}
P_\text{res}(R>R_c) =  \int_{R_c}^{\infty} R^{d-1} dR \frac{|T_{\alpha, \eta}(R)|}{\delta}
\end{equation}
where $\delta \sim E_{\alpha}^{\nu d}$ is the typical level spacing at energy $E_{\alpha}$. 
Thus, we conclude that the probability of hopping resonances at a distance $R$ behaves as 
\begin{equation}
P_\text{res}(R>R_c) \sim V^2 |E_{\alpha}|^{-1} R_c^{-(d/2 + 3/(2\nu)}
\end{equation}
For any non-zero $\epsilon$ and any non-zero $E_{\alpha}$ there is an $R_c$ such that $P_\text{res}(R>R_c) < \epsilon$. Thus, the effective long range hopping falls off sufficiently rapidly that the probability of having long range hopping resonances vanishes, and thus the effective long range hopping does not present an obstacle to the construction of marginally localized integrals of motion.

\begin{figure}[bbb]
\includegraphics[width = 1.5in]{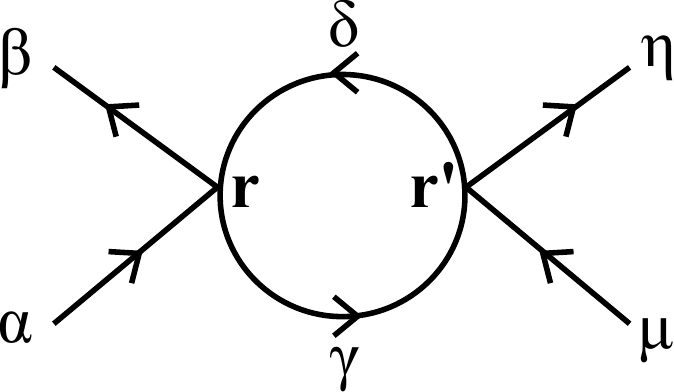}
\caption{Diagram illustrating flip-flop process}
\label{fig:LowOrderFF}
\end{figure}

\subsection{Flip-flop resonances}
Next, we consider the matrix elements of $\hat T$ between two particle states, where the particles in question are separated by a distance $R$. Such processes may be represented diagrammatically by processes of the form shown in Fig.~\ref{fig:LowOrderFF}, and may be interpreted as `flip flop' processes which can transport energy over long distances \cite{pairhop}. The effective matrix element for this sort of `flip-flop' process scales as 

\begin{widetext}
\begin{align}
T_{\alpha \mu, \beta \eta}(R) &= \sum_{ \gamma \delta}
 \frac{\lambda_{\alpha\beta\gamma\delta}\lambda_{\gamma\delta\mu\eta}} 
{(E_{\alpha} + E_{\beta}) - (E_{\gamma} + E_{\delta})}\\
&= \frac{V^2}{(\xi_{\alpha} \xi_{\beta} \xi_{\mu} \xi_{\eta})^{d/2}} \sum_{ \gamma \delta} \frac{\exp\left[-R\left(\frac{1}{\xi_{\gamma}} +\frac{1}{\xi_{\delta}}\right)\right]}{\left( \xi_{\gamma}^d \xi_{\delta}^d\right)\[\(E_{\alpha} - E_{\beta}\) - \(E_{\gamma} - E_{\delta}\)\]} \exp \left(-i \phi_{\gamma \delta}\right) 
\end{align}
\end{widetext}
where again $\phi_{\gamma \delta}$ is a random phase. The arguments proceed exactly as for the hopping resonances (except that there is one fewer internal variable to sum over), and lead to the result 
\begin{equation}
\langle |T_{\alpha \mu, \beta  \eta}(R)|\rangle = V^2 \frac{|E_{\alpha} E_{\beta}E_\mu E_\eta|^{d \nu/2}}{|E_\alpha-E_\beta|} R^{-(d+1/\nu)}
\end{equation}
Since the strongest contributions will come from states $\alpha,\beta,\mu,\eta$ with similar energies, we take $E_{\gamma,\delta,\beta}\approx E_\alpha$ in the numerator, and replace the energy denominator, $|E_\alpha-E_\beta|$ by minimum level spacing for states of size $\xi(E_\alpha)\approx E_\alpha^{\nu d}$, producing: 
\begin{equation}
\langle |T_{\alpha \mu, \beta \eta}(R) |\rangle = V^2 E_\alpha^{\nu d} R^{-(d+1/\nu)}
\end{equation}

We now ask what is the probability that a {\it particular} typical transition has a resonant pair within a distance $R$. The typical SP level spacing at an energy $E_{\alpha}$ is $\sim \xi_{\alpha}^{-d} \sim E_{\alpha}^{\nu d}$. The number of transitions which could be paired with the transition of interest goes as $R^d$. Thus, the minimum energy change for flip-flops involving a particular typical transition goes as $E_{\alpha}^{\nu d}R^{-d}$. Meanwhile, the matrix element falls off as $V^2 E_{\alpha}^{\nu d}R^{-(d+1/\nu)}$. The probability that a particular transition has a resonant partner on the length scale $R$ is thus $\sim V^2 R^{-1/\nu}$, which goes to zero as $R \rightarrow \infty$ for any $\nu$. 
Thus, the probability of having a flip-flop resonance will drop below $\epsilon$ on length scales $R > R_c$, where $R_c = (V^2/\epsilon)^{\nu}$. It is  instructive to contrast with the usual MBL case, Fig.1a, where the SP localization length is finite at all energies. This situation may be modeled by taking $\nu \rightarrow 0$, so that all states (except the exact zero energy state) are localized with localization length of order unity. Taking this limit predicts that for the traditional MBL problem, the length scale $R_c$ beyond which the probability of flip-flop resonances drops below $\epsilon$ should grow slower than any power law of $\epsilon$. Meanwhile, the situation with a mobility edge can be modeled by taking $\nu \rightarrow \infty$, such that all states with $E<1$ are delocalized. In this case there will be no finite length scale $R_c$ beyond which the probability of flip flop resonances drops below $V^2$, indicating instability of localization to interactions at finite energy density in partially filled bands with a SP mobility edge. 

\subsection{Flip-flop assisted hopping}
Finally, we consider the matrix elements of $\hat T$ between three particle states. 
\begin{figure}
\includegraphics[width = 2.3in]{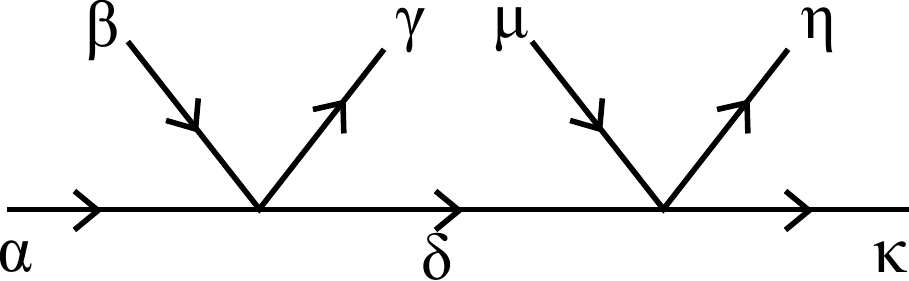}
\caption{The flip-flop assisted hopping process \label{assist}}
\end{figure}
This process is represented diagrammatically by Fig.\ref{assist}, and may be interpreted as a `flip-flop assisted hopping,' whereby a particle hops from $\vec{r_1}$ to $\vec{r_2}$ by triggering SP transitions at both $\vec{r_1}$ and $\vec{r_2}$. We are interested as before in determining how this matrix element scales with $R = |\vec{r_1} - \vec{r_2}|$. This process has matrix element
\begin{widetext}
\begin{align}
T_{\alpha \beta \mu, \gamma \eta \kappa}(R) &= \sum_{ \delta}
 \frac{\lambda_{\alpha\beta\gamma\delta}\lambda_{\delta\mu\eta \kappa}} 
{(E_{\alpha} + E_{\beta}) - (E_{\gamma} + E_{\delta})}\\
&= \frac{V^2}{(\xi_{\alpha} \xi_{\beta} \xi_{\gamma} \xi_{\mu} \xi_{\eta}\xi_{\kappa})^{d/2}} \sum_{ \delta} \frac{\exp\left[-\frac{R}{\xi_{\delta}}\right]}{\xi_{\delta}^d\[\(E_{\alpha} - E_{\beta}\) - \(E_{\gamma} - E_{\delta}\)\]} \exp \left(-i \phi_{ \delta}\right) 
\end{align}
\end{widetext}
where again $\phi_{\delta}$ is a random phase. The arguments proceed exactly as for the flip flop resonances (except that there is one fewer internal variable to sum over), and lead to the result 
\begin{equation}
\langle |T_{\alpha \beta  \mu, \gamma \eta \kappa}(R)|\rangle = V^2 \frac{|E_{\alpha} E_{\beta}E_\gamma E_\mu E_\eta E_\kappa|^{d \nu/2}}{|E_\alpha-E_\beta|} R^{-\frac12(d+1/\nu)}
\end{equation}
Since the strongest contributions will come from states $\alpha,\beta,\mu,\eta$ with similar energies, we take $E_{\gamma,\delta,\beta}\approx E_\alpha$ in the numerator, and replace the energy denominator, $|E_\alpha-E_\beta|$ by minimum level spacing for states of size $\xi(E_\alpha)\approx E_\alpha^{\nu d}$, producing: 
\begin{equation}
\langle |T_{\alpha \beta  \mu, \gamma \eta \kappa}(R)|\rangle = V^2 E_\alpha^{2\nu d} R^{-\frac{1}{2}(d+1/\nu)}
\end{equation}

There are $\sim R^d$ sites within a distance $R$ of $\psi_{\alpha}$ to which a particle could hop via this `flip-flop assisted' hopping process. Thus, the minimum level spacing for such a process falls off as $R^{-d}$. Meanwhile, the matrix element for this process falls off only as $R^{-\frac{1}{2}(d+1/\nu)}$. Thus, the probability that there is no `assisted hopping' resonance at a length scale $R$ vanishes at long distances as a power law of $R$ for $\nu d > 1$ (for $\nu d = 1$ it vanishes logarithmically with $R$). Thus, for $\nu d \ge 1$, a particle can always find a (distant) resonant site to hop to, with some assistance from flip flops. This establishes that localization is impossible in the presence of interactions if 
\begin{equation}
d \nu \ge 1 \label{result}
\end{equation}
This is the main result of this section, and concludes our discussion of the matrix elements of $\hat T$ at $O(V^2)$ (we note that $\hat T$ has no matrix elements between four particle states at $O(V^2)$ because $\hat T$ must be represented by a connected diagram). Next, we discuss whether we can get a more stringent condition for breakdown of localization at higher orders in $V$. 

\subsection{Higher orders in $V$}
\begin{figure}
a)\includegraphics[height = 1.25in]{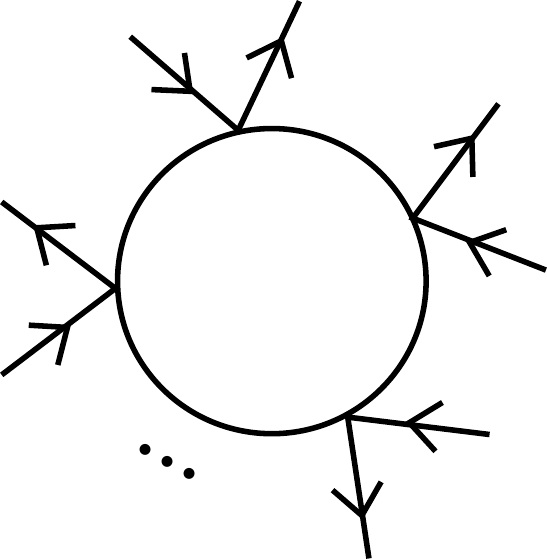}\hspace{.2in}
b) \includegraphics[height = 1in]{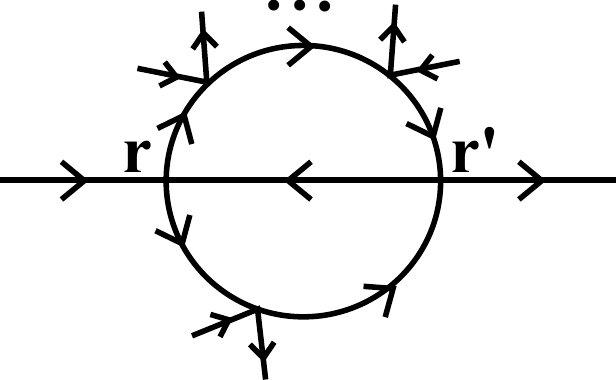}
\caption{ Some high order diagrams contributing to $\hat T$}
\label{fig:HighOrderFF}
\end{figure}

We now discuss whether a more stringent condition on breakdown of localization could be obtained by going to higher order in $V$ and considering diagrams, e.g. of the sort illustrated in Fig.\ref{fig:HighOrderFF}.
 We will conclude that the answer is no, and that (\ref{result}) is the criterion for perturbative breakdown of marginal localization. We begin by noting that the diagrammatic representation of the Born series (\ref{Born}) contains vertices, internal lines (which begin and end on a vertex), and external lines (which are connected to vertices only at one end). At $O(V^n)$ in perturbation theory, we will have $n$ vertices, $m$ internal lines and $p$ external lines, where $2 m + p = 4 n$. Since only connected diagrams contribute to the $\hat T$ matrix, every vertex must be connected to at least one internal line, and all but two of the vertices must be connected to at least two internal lines. This gives the constraint $  m \ge (n-1)$.

Now, the $R$ scaling depends only on the number of internal lines $m$. After summing over internal variables in a rms sense, we find that the $R$ scaling has the form
\begin{equation}
T(R) \sim \sqrt{\left[\int^{R^{-1/\nu}} dE E^{-\nu d} (E^{\nu d})^2\right]^m} 
\end{equation}
where the first factor $E^{-\nu d}$ comes about when we change the sum over states to an integral, and the second factor $(E^{\nu d})^2$ comes from the normalization factors associated with the internal lines. This gives 
\begin{equation}
T(R) \sim R^{-\frac{m}{2} (d+1/\nu)}
\end{equation}

Now, if we evaluate the matrix element of $\hat T$ that connects a particle at a given position to particles at $\tilde n$ other positions within a ball of radius $R$ (i.e. if we look at diagrams containing $n \ge \tilde n+1$ vertices that are connected to at least one external line), then the minimum level spacing for such processes will fall off as $\sim R^{-\tilde n d}$. If this level spacing falls off faster than the corresponding matrix element, then the probability of having resonances will approach one at large $R$, so that localization will be unstable. The condition for delocalization thus takes the form
\begin{equation}
\tilde n d > \frac{m}{2}\left(d+\frac{1}{\nu}\right)
\end{equation}
Thus, to maximize our chances of delocalization, we should maximize $\tilde n$ and should minimize $m$, subject to the contraints we have already identified $\tilde n +1 \le n$ and $m \ge n-1$. Substituting $\tilde n = n-1$ and $m = n-1$ into the above formula, we find that the condition for delocalization becomes 
\begin{equation}
(n -1) d > \frac{n-1}{2}\left(d+\frac{1}{\nu}\right)\Rightarrow d \nu > 1
\end{equation}
Thus, the tightest constraint that we could possibly get within perturbation theory is the constraint (\ref{result}) coming from the flip-flop mediated hopping process i.e. delocalization occurs if $d \nu > 1$. Meanwhile, for $d \nu < 1$, the probability of resonances vanishes at large $R$ at all orders in perturbation theory, so that marginal localization appears to be perturbatively stable. For $d \nu = 1$ the probability of having a resonance, calculated in the manner of (13), grows logarithmically with $R$, suggesting that the system should be delocalized (this is also consistent with results on non-interacting models with critically long range hopping \cite{Levitov2}). 

\subsection{Systems with a vanishing critical density of states}
The result (\ref{result}) is slightly modified if the density of states vanishes as $E^{\Upsilon}$ in the vicinity of the critical energy. In this case, the integrals over energy come with a factor of $E^{\Upsilon}$, and the matrix elements with $m$ internal lines are of order 
\begin{equation}
T(R) \sim R^{-\frac{m}{2} (d+(1+ \Upsilon)/\nu)}
\end{equation}
In this case similar arguments establish that delocalization occurs if 
\begin{equation}
\nu d \ge 1 + \Upsilon
\end{equation}
The situation where the density of states diverges at the critical energy can also be treated within the above formalism, by taking $\Upsilon < 0$. 

\subsection{An intuitive explanation for these results}
We now draw attention to a very simple and intuitive explanation for the delocalization criterion (\ref{result}). On any length scale $R$, there are states with energies $|E| < R^{-1/\nu}$ that look effectively extended. Meanwhile, there are $\xi^d(E)$ states at an energy $E$ that have support at a particular position $\vec{r}$. Thus, the number of states to which we can couple with a local interaction which are extended on the scale $R$ is 
\begin{equation}
N(R) \sim \int_0^{R^{-1/\nu}} dE E^{\Upsilon} \xi^d(E) = \int_0^{R^{-1/\nu}} dE E^{\Upsilon - \nu d}
\end{equation}
If $\nu d < 1+ \Upsilon$ then the `bath' of extended states to which we can locally couple contains a finite number of states (and the number of states in the bath goes at zero as the length scale $R\rightarrow \infty$). In this case, localization can be perturbatively stable. In contrast, if $\nu d > 1+\Upsilon$, then the above integral diverges near $E=0$, such that the `bath' to which we can locally couple contains an infinite number of extended states at any length scale $R$. In this case, the algebraically delocalized states near the critical energy form a good bath, and are able to thermalize the system by mediating long range interactions that fall off sufficiently slowly with $R$, as we have discussed. 

\subsection{Characteristic length- and time- scales for delocalization }
We now estimate the length scale and timescale for the resonances that give rise to a breakdown of localization. Recall that the $T$ matrix elements between states separated by a distance $R$ at order $n+1$ in perturbation theory scale as $V^{n+1} R^{-n(d+1/\nu)/2}$, while the level spacing  scales as $R^{-n d}$. The length scale $R_c(n)$ on which resonances appear at order $n+1$ in perturbation theory is the length scale at which the matrix elements and the level spacings become comparable i.e. $V^{n+1} R_c^{-n(d+1/\nu)/2} = R_c^{-nd}$. This happens at a critical length scale $R_c$, where
\begin{equation}
R_c(n) \sim V^{-\frac{2\nu}{d \nu -1} \frac{n+1}{n}} \label{eq:rcn}
\end{equation}
By inspection, the smallest $R_c$ comes from $n \rightarrow \infty$ for which
\begin{equation}
R_c \sim V^{-2 \nu/(d \nu -1)} \label{Rc}.
\end{equation}
The scaling of $R_c$ in Eq.~\ref{Rc} controls, for example, the length scale at which the longitudinal conductivity ceases to follow (\ref{eq:DCConductivity}) and saturates instead to a non-zero constant. Remarkably, even though breakdown of perturbation theory occurs already at second order in perturbation theory, the shortest length scale on which resonances appear is controlled by high orders of perturbation theory. 

We can also estimate the timescale on which breakdown of localization occurs. Processes at order $n+1$ cause a breakdown on length scales $R_c(n)$, given by (\ref{eq:rcn}). The matrix element for these processes is of order $V^{n+1} R_c(n)^{-n(d+1/\nu)/2} = R_c(n)^{-nd}$, and the level spacing is $R_c(n)^{-nd}$, so that the density of final states is $R_c(n)^d$. A straightforward application of the Golden Rule implies that the timescale on which breakdown of localization happens due to processes at $n+1$ order in perturbation theory is 
\begin{equation}
t_c(n) \sim R_c(n)^{nd} = V^{- \frac{2 \nu d (n+1)}{\nu d -1}}\label{Tcn}
\end{equation}
Thus we see that the shortest timescale on which breakdown occurs is controlled by $n=1$ (i.e. by processes at second order in perturbation theory), and is of order
\begin{equation}
t_c \sim V^{- \frac{4 \nu d}{\nu d -1}}\label{eq: Tc} 
\end{equation}
For the case of integer quantum Hall, with $\nu = 2.3$ and $d=2$, $R_c \sim V^{-1.3}$ and $t_c \sim V^{-2.5}$. We note that the crossover timescales $R_c$ and $t_c$ both diverge as $V\rightarrow 0$, recovering marginal localization in the non-interacting limit ($V=0$). 

So far we have suppressed temperature-dependent occupation factors in estimating the T-matrix amplitudes. The breakdown of marginal localization happens (for a closed quantum system) because the `internal heat bath' presented by the states with diverging localization length allows the system to thermalize. At low temperature, the coupling to this internal bath is suppressed by a factor of $\exp(-\Delta/T)$, where $\Delta$ is the detuning of the chemical potential from the critical energy - i.e. the crossover scales $R_c$ and $t_c$ will be enhanced by an `Arrhenius' factor $\sim \exp(c \Delta/T)$, where $c$ is a positive numerical pre factor that depends on $\nu$ and $d$. Thus, the crossover scales will also diverge in the zero temperature limit, so that the zero temperature system will indeed be localized, as long as the Fermi level is not at the critical energy. 

The emergence of a new length scale and timescale in the interacting problem suggests that the scaling behavior near the zero temperature quantum critical point \cite{QuantumHallPlateau} may be richer for the interacting system than for the non-interacting system. In particular, the presence of an additional interaction related length scale $R_c$ may help explain why the experimentally observed longitudinal conductance is not temperature independent at the critical point \cite{QuantumHallPlateau} - the longitudinal conductance could depend on the ratio of $R_c$ to the thermal length $\hbar / kT$. Additionally, if one examines the scaling of e.g. the longitudinal conductance with temperature at a non-zero detuning $\Delta$ from the critical point, then the coupling to the `internal' heat bath will provide a mechanism for a thermally activated scaling of the conductance, with an activation gap proportional to the detuning from the critical point. Of course, any real experimental system will also be coupled to an `external' heat bath (e.g phonons), which will allow for variable range hopping at any non-zero temperature, and this variable range hopping contribution to the conductance will dominate over the `activated' contribution at the lowest temperatures. However, if one observed a regime where the low temperature longitudinal conductance was activated, with an activation gap proportional to the detuning from criticality, then this may be tentatively identified as evidence for the `delocalization due to the internal bath' discussed in this paper. An additional complication is that we have considered short range interactions, while in real quantum Hall systems there are also Coulomb interactions (unless screened by a nearby metallic gate).  The inclusion of long-range Coulomb interactions may introduce additional corrections to scaling that are not captured by the present theory.

\subsection{Multifractality}
Thus far we have assumed that the wave functions were uniform on length scales short compared to $\xi$. In fact, this is not the case, because close to criticality the wave functions exhibit multi fractal behavior on length scales less than $\xi$. It is known \cite{Chalker, EversMirlin} that for wave functions close to criticality, the disorder averaged two point correlations fall off as $\langle \psi_{\alpha}^*(0) \psi_{\beta}(0) \psi_{\beta}^*(r) \psi_{\alpha}(r) \rangle \sim \frac{1}{\xi^{2d}} (\xi/r)^{\eta}$ for $r< \xi$, where $\eta>0$ is a multi fractal exponent. This suggests that the wave functions actually have higher amplitudes on scales $r \ll \xi$ than would be expected based on the `plain vanilla' model we were working with thus far. This will tend to {\it enhance} the matrix elements of the $\hat T$ matrix over and above the estimates above. Thus, the condition $\nu d > 1 + \Upsilon$ may be viewed as a lower bound condition for delocalization. If we take into account the effects of multifractality in a simple minded way by attaching a factor of $\left(\xi_{int}(E)/R\right)^{\eta/2}$ to each internal line, then we will conclude that the contribution to the $\hat T$ matrix from a diagram with $m$ internal lines (after rms summation over internal variables) will be 
\begin{eqnarray}
T(R) &\sim& \sqrt{\left[\int^{R^{-1/\nu}} dE E^{\Upsilon-\nu d} (E^{\nu d})^2 \left(\frac{E^{-\nu}}{R}\right)^{\eta}\right]^m} \nonumber\\
&\sim& R^{-\frac{m}{2}\left(d+\frac{1+\Upsilon}{\nu}\right)}
\end{eqnarray}
which is independent of $\eta$ - i.e. if we take multifractality into account in this simple minded way then the threshold for delocalization is unchanged, and is  $\nu d \ge 1 + \Upsilon$. A more sophisticated treatment of multi-fractality may reduce the critical value of $\nu$ somewhat, but (as we have discussed), multifractality can only make it {\it easier} to get delocalization.

\section{Discussion and Conclusion}
We have discussed marginally localized systems, which for partially filled bands at non-zero energy density are characterized in the non-interacting limit by sub diffusive relaxation (\ref{eq:Subdiffusion}), a scale dependent DC conductivity that vanishes in the thermodynamic limit (\ref{eq:DCConductivity}) and an entanglement entropy of the excited eigenstates that is intermediate between area law and volume law (\ref{eq:entanglement}). We have shown that such systems are generically unstable to interactions, since interactions trigger a breakdown of localization for any $\nu > 1/d$. We further note that, for most systems, general arguments (\cite{Harris, Chayes} and Appendix) place a lower bound, $\nu \ge 2/d$, on the localization length exponent. Thus, we have shown that marginally localization is generically unstable to interactions, for partially filled bands at non-zero energy density.

We therefore predict that in a marginally localized system such as integer quantum Hall, the inclusion of arbitrarily weak short range interactions will cause the entanglement entropy in the excited many body eigenstates to become volume law, instead of the fractal scaling shown in (\ref{eq:entanglement}).  We also predict that in two dimensional weakly interacting marginally localized systems, the scale dependent conductivity at nonzero energy density will follow (\ref{eq:DCConductivity}) on small length scales, but will saturate to a non-zero constant on large length scales.  Similarly, relaxation in such marginally localized systems should be sub-diffusive according to (\ref{eq:Subdiffusion}) for short times, but should cross over to diffusive relaxation on long timescales. We have estimated the crossover length scales and timescales, and have also discussed in Sec.IV-G the interaction induced corrections to scaling in the vicinity of the zero temperature quantum critical point. These ideas may be directly tested in experiments in cold-atomic systems and possibly also in solid-state materials for which the electron-phonon interactions are negligible, so that the electronic system well approximates an isolated quantum system. We speculate in particular that disordered graphene in the quantum Hall regime may provide a playground to test these ideas, due to its weak electron-phonon coupling and isolation from any substrate in suspended devices.

While our arguments were for short range interactions, the question of whether MBL can survive with long range interactions has also been recently studied \cite{Yao}. It was claimed there that full MBL is perturbatively stable against weak interactions as long as the interactions fall off with distance faster than $1/r^p$, where $p$ is a critical power that depends on the model in question. For interactions that fall off faster than $1/r^p$, our arguments can be straightforwardly generalized to demonstrate that marginal localization is perturbatively unstable. The situation with truly long range interactions (which fall off slower than $1/r^p$), is beyond the scope of the present work. 

We note that there are three classes of short range interacting system that may evade our arguments. If the divergence in the localization length is accompanied by a density of states (DOS) which vanishes as $E^{\Upsilon}$, our argument only predicts delocalization for $d \nu > 1+ \Upsilon$. Given a sufficiently rapid vanishing of the DOS at the critical energy, marginal localization could be perturbatively stable. Additionally, in systems where the critical energy is pinned (e.g. by symmetry) to a particular value and is locally insensitive to disorder, the Harris criterion does not apply and we can have $\nu < 2/d$ (see \cite{Singh} and Appendix). A particular example of this is models of particle-hole symmetric localization in one dimension \cite{Motrunich}, where the divergence in the localization length is only a logarithmic function of energy. In this case our analysis would suggest that marginal localization could be perturbatively stable. Finally, our arguments establish that marginal localization is unstable for $\nu > 1/d$, where $\nu$ is the critical exponent for the localization length of {\it typical} states. However, the Harris/Chayes criteria only constrain the critical exponent for the {\it mean} localization length $\nu_{mean} \ge 2/d$. For infinite randomness critical points, $\nu_{typical}$ and $\nu_{mean}$ can be markedly different\cite{Fisher}. Thus, it is conceivable that there may be systems with infinite randomness critical points where $\nu_{mean} \ge 2/d$ (satisfying Harris/Chayes), but $\nu_{typical} < 1/d$, so that marginal localization could be perturbatively stable. Investigations of whether such marginally localized systems can be stable to interactions would be a fruitful topic for future work.

 \section{Acknowledgements}
We thank David A. Huse,  Siddharth Parameswaran, S.L. Sondhi, Romain Vasseur, Ashvin Vishwanath, Ravin Bhatt, and Norman Y. Yao for helpful conversations. We also thank David A. Huse and Sarang Gopalakrishnan for a critical reading of the manuscript. RN is supported by a PCTS fellowship.

 \appendix
\section{Harris/Chayes bounds for Localization Length Exponent \label{app:Harris}}
The generating function for two-body Green's functions with fixed energy $\e$ of a non-interacting quantum system in $d$ spatial dimensions can be written in terms of a partition function with a $d$-dimensional action:
\begin{align} Z(\e) = \int D[\bar{\psi},\psi] e^{i\bar{\psi}(\e+i0^+-H)\psi} \end{align}
where $H$ is the single-particle Hamiltonian including random disorder.  A diverging SP localization length corresponds to a critical value of $\e_c$ for which $Z(\e)$ exhibits a phase transition.  Harris's argument, which we briefly review below, can then be applied to this $d$-dimensional theory to constrain the correlation length exponent $\nu$. 

Within a region of size $L$, the disorder potential gives a random shift to the average value of $\e$ within the region of order $\delta(L) \sim L^{-d/2}$.  Then, in order to determine the average value of $\e-\e_c$ to accuracy $\delta$, one needs to look at lengthscale $L\sim \delta^{-2/d}$.  On the other hand, if the transition occurs inside the single particle band, one should be able to determine on which side of the transition the system lies by examining a sub-system with size $L$ much bigger than the correlation length $\xi(\e)\sim |\e-\e_c|^{-\nu}$.  Together these considerations require that the random shift in average $\e$ in a subsystem of size $\xi(\e)$ goes to zero no slower than than the distance to the critical point  $\displaystyle \lim_{\e\rightarrow \e_c} \delta\(\xi(\e)\)/|\e-\e_c|=0$, or:
\begin{align} d\nu \ge 2 \label{eq:Harris},\end{align}
the celebrated Harris criterion. 

This argument applies only if the critical energy can locally be shifted by disorder \cite{Singh}. In situations where the extended states are pinned at a particular energy independent of disorder, perhaps by a symmetry, then the Harris bound does not apply, and the system can have $\nu d < 2$. An example of this situation is the case of particle-hole symmetric localization \cite{Motrunich}, where the delocalized states sit exactly at $E=0$ because of particle hole symmetry independent of the strength of the (random hopping) disorder. In this case, the divergence of the localization length is only a logarithmic function of the energy (i.e. slower than any power law), and our argument suggests that marginal localization could be at least perturbatively stable.


\begin{thebibliography}{99}
\bibitem{pairhop}
Permuting the arrows in the diagram also gives a `long range pair hopping' process whereby two particles at energies $\pm E$ and position $\vec{r_1}$ hop to two states at energy $\approx \pm E$ and position $\vec{r_2}$ with $|\vec{r_2} - \vec{r_1}| = R$. However, this matrix element scales with $R$ in the same way as the flip flop process, and so we do not discuss it separately.
\end{thebibliography}
 \end{document}